\documentstyle[epsfig]{aipproc}

\begin{document}

\title{GRB Redshift Distribution \\ is Consistent with GRB Origin \\
in Evolved Galactic Nuclei}

\author{V.I. Dokuchaev$^1$, Yu.N. Eroshenko$^1$,
and L.M. Ozernoy$^{2,3}$}
\address{$^1$Institute for Nuclear Research, Russian Academy of Sciences,
Moscow, 117312, Russia\\
\noindent $^{2}$George Mason University, Fairfax, VA 22030-4444, USA\\
\noindent $^{3}$Goddard Space Flight Center, Code 685, Greenbelt, MD 20771,
USA}

\maketitle

\begin{abstract}
Recently we have elaborated a new cosmological model of gamma-ray
burst (GRB) origin \cite{we2}, which employs the dynamical
evolution of central dense stellar clusters in the  galactic
nuclei. Those clusters inevitably contain a large fraction of
compact stellar remnants (CSRs), such as neutron stars (NSs) and
stellar mass black holes (BHs), and close encounters between them
result in radiative captures into short-living binaries, with
subsequent merging of the components, thereby producing GRBs
(typically at large distances from the nucleus).

In the present paper, we calculate the redshift distribution of the
rate of GRBs produced by close encounters of NSs in distant
galactic nuclei.
To this end, the following steps are undertaken: (i) we establish a
connection between the parameters of the fast evolving central stellar
clusters ({\it i.e.} those
for which the time of dynamical evolution exceeds the age of the Universe)
 with masses of the forming central supermassive black holes
(SMBHs) using a dynamical evolution model; (ii) we connect
these masses with the inferred mass distributions of
SMBHs in the galactic nuclei and the redshift distribution of quasars by
assuming a certain `Eddington luminosity phase' in their activity;
(iii) we incorporate available observational data on
the redshift distribution of quasars as well as  a recently found
correlation between the masses of galaxies and their central SMBHs.
The resulting redshift distribution of the GRB rate, which accounts
for both fast and slowly evolving
galactic nuclei is consistent with that inferred from the BATSE
data if the fraction of fast evolving galactic nuclei is in the range
$\varepsilon\simeq 0.016-0.16$.
\end{abstract}

\section*{Introduction}

In our previous work \cite{we2}, we have considered the generation of
GRBs due to the radiative collisions of CSRs (for simplicity, taken to be
solely NSs) in the central stellar clusters of evolved galactic nuclei.
Depending on its initial radius and mass, the characteristic dynamical
evolution time of a cluster,
$t_{e}$, can either exceed the age of the Universe, $t_0$,
(we call such clusters as `{\it slowly evolving clusters}'
hereinafter) or be less than $t_0$ (`{\it fast evolving clusters}').
The GRB rate from slowly evolving clusters does not depend
on redshift of the host galactic nuclei. Following an approach
developed in
\cite{qs87}, the mean rate of radiative collisions of NSs in the
fiducial galactic nucleus with mass of a central cluster $M$,
radius $R$, and stellar velocity dispersion $v=(GM/2R)^{1/2}$ is given by
\cite{we2}:
\begin{equation}
\dot N_{c}\simeq9\sqrt2{\left(\frac{v}{c}\right)}^{17/7}
\frac{c}{R}\simeq 5.8\cdot10^{-6}\,
{\left(\frac{M}{10^7\,{\rm M}_{\odot}}\right)}^{17/14}
{\left(\frac{R}{0.1\,{\rm pc}}\right)}^{-31/14}
\frac{\mbox{events}}{\mbox{yr~galaxy}}~,
\label{Nrate}
\end{equation}
where we use for normalization the likely values of $M$ and $R$
to fit the inferred rate of GRBs. The GRB rate from a unit
comoving volume is $\dot n_{0}=\langle \dot N_{c} \rangle n_{g}$,
where the number density of galaxies $n_g\sim10^{-2}$~Mpc$^{-3}$
and $\langle\ldots\rangle$ is an averaging throughout all types of
slowly evolving galactic nuclei.

A galactic
nucleus with the gravitationally dominating SMBH of mass $M_h>M$
might belong to slowly evolving galactic nuclei as well.
The corresponding GRB rate from such a nucleus is given by \cite{we2}
\begin{equation} \label{NBHrate}
\dot
N_{c,h}\simeq\frac{9}{2\sqrt2}{\left(\frac{Nm}{M_{h}}\right)}^{2}
\left(\frac{v}{c}\right)^{17/7}\frac{c}{R}~,
\end{equation}
which differs from that given by Eq.~(\ref{Nrate}) by a
factor ${\left({M}/{M_{h}}\right)}^{2}\ll 1$.

One more possible contribution to GRB rate from a
slowly evolving system is due to
the coalescence of tight NS binaries in the
galactic discs (which is a commonly believed scheme of cosmological GRB
origin). In this paper, we assume that there was no strong evolution
with cosmological time of the number of tight NS binaries
both in the galactic disks and in the slowly evolving nuclei so
that all these contributions result in the
redshift-independent GRB rate $\dot n_0
\simeq10^{-8}$~yrs$^{-1}$~Mpc$^{-3}$.

Meanwhile the actual parameters of galactic nuclei vary in a wide
range and some central clusters may, in fact, be
fast evolving ($t_e<t_0$). The GRB rate from  the fast evolving
clusters depends on redshift. Below, we find the
redshift distribution of the GRB rate from both slow and fast evolving
clusters by (i) modelling the dynamical evolution of a fast
evolving cluster in the galactic nucleus, which leads to the production of
a SMBH; (ii) connecting the parameters of a fast evolving
cluster with the mass of its central SMBH, and (iii) using the observed
distributions in luminosity and redshift for quasars, as well as in mass for
SMBHs in the galactic nuclei.

\section*{Fast Evolving Galactic Nuclei}

We introduce a dimensionless time variable $y\equiv t_0/t_e=
(1+z_e)^{3/2}$, where $t_e$ is the time necessary for a SMBH formation,
$z_e$ is the redshift at that instant. Evidently, $y>1$
for  clusters collapsed to the present epoch $t=t_0$ (fast
evolving clusters) and $y<1$ for clusters which do not
collapse by $t_0$ (slowly evolving clusters). As we show elsewhere
\cite{deo99}, the dynamical evolution of a stellar cluster,
in the framework of a homologous Fokker-Plank
approximation \cite{qs87}, leads to the following relationship between
the initial mass of the cluster and the mass of the central SMBH
formed there:
\begin{equation}
M_f=2\cdot10^6y^{2.22}\left[
\frac{M_i}{2.8\cdot10^8M_{\odot}}
\right]^{5.44}M_{\odot}.
\label{mfmi}
\end{equation}
There are serious observational indications \cite{9712076},
\cite{0006}, \cite{0003} that a fraction of galaxies with a
central SMBH might be as high as  $\varepsilon\sim 0.1$ and that
there is a correlation between the central SMBH mass, $M_h$, and
the luminosity of the host galaxy bulge, $L_s$, which in a
simplified form can be expressed as
$M_h\simeq10^{-2{.}5}(L_s/L_{\odot})M_{\odot}$. This correlation,
taken together with the Schechter luminosity function of galaxies
\cite{sheh}, gives the mass distribution of SMBHs,
$\phi_1(M_h)dM_h$ with $\phi_1(M_h)\propto\varepsilon$. Then
we use the observed \cite{boyle91} distribution of quasars
in absolute magnitude $M_B$ and redshift $z$, $\phi_2(M_B,z)dM_Bdz$, 
for $z\leq3$. Finally, we assume the existence of an
`Eddington luminosity phase' in the evolution of quasars, when
they shine with the Eddington luminosity (or with its fraction
$\lambda\leq 1$) during a certain finite time interval, which
duration only depends upon $\lambda$  and the initial SMBH
mass. In our calculations, the `final' (i.e., by the end of the
dynamical evolution) mass of a galactic nucleus is the initial
mass of its new-born SMBH. The latter then experiences the
`Eddington luminosity phase', during which its mass grows up to
the SMBH mass in the actually observed quasar. As a result, we
find the mass distribution of {\it active} SMBHs,
$P(M_f,y)dM_fdy$, in terms of the variables $y$ and mass $M_f$ at
the instant of the SMBH formation (in the limit
$M_f<3\cdot10^6M_{\odot}$). Thus, these active SMBHs arise
due to the dynamical evolution of galactic nuclei with fast
evolving central clusters. 

\section*{Distribution of GRB Rate in Redshift}

In fast evolving galactic nuclei, the resulting rate of GRB generation per
unit of comoving volume is  given as a function of redshift $z$ by
\begin{equation}
\dot n_{ev}(z)=\int
dM_f\int\limits^{x(z)^3}_1P(M_f,y)\dot N_{ci}(M_i(M_f),y)
\frac{dy x^{3\xi}} {(x^3-y)^{\xi}}~,
\label{dotnz}
\end{equation}
where $x=(1+z)^{1/2}$, $\xi\simeq0.6$, $\dot N_{ci}$ is found
from Eq.~(\ref{Nrate}) with $M=M_i$ and $R=R_i$, and function
$M_i(M_f)$ is the inverse function given by Eq.~(\ref{mfmi}). Numerical
integration over the variable $M_f$ in Eq.~(\ref{dotnz}) proceeds in the
range $M_f\sim10^5~-~3\cdot10^6\,M_{\odot}$, which
approximately correspond to the range of
$M_i\sim10^7~-~2\cdot10^8\,M_{\odot}$.

The total rate of GRBs from {\it all} kinds
of galactic nuclei, $\dot n(z)$, 
includes also the contribution $\dot
n_0\simeq 10^{-8}$~Mpc$^{-3}$~yr$^{-1}$ from all slowly evolving
galactic nuclei and disks with $t_e>t_0$,
i.e. $\dot n(z)=\dot n_0 + \dot n_{ev}(z)$.
The total GRB rate as a function of $z$ can be fitted 
by a one-parametric function: $\dot n(z)=\dot n_0(1+z)^\beta$,
where the range of $\beta$, $1{.}5<\beta<2$, accounts for
the statistical uncertainties \cite{horack}. This function
is shown in Fig.~1. 
The consistency between the calculated 
and inferred GRB rates is achieved 
at  $\varepsilon/\lambda\simeq 0.16$.  
Since quasars shine at $L\simeq (0.1-1)~L_{\rm Ed}$ (e.g., \cite{wandel}),
it implies that $\varepsilon\simeq 0.016-0.16$, which is consistent with
available data [4]-[6].

\begin{figure} 
\centerline{\epsfig{file=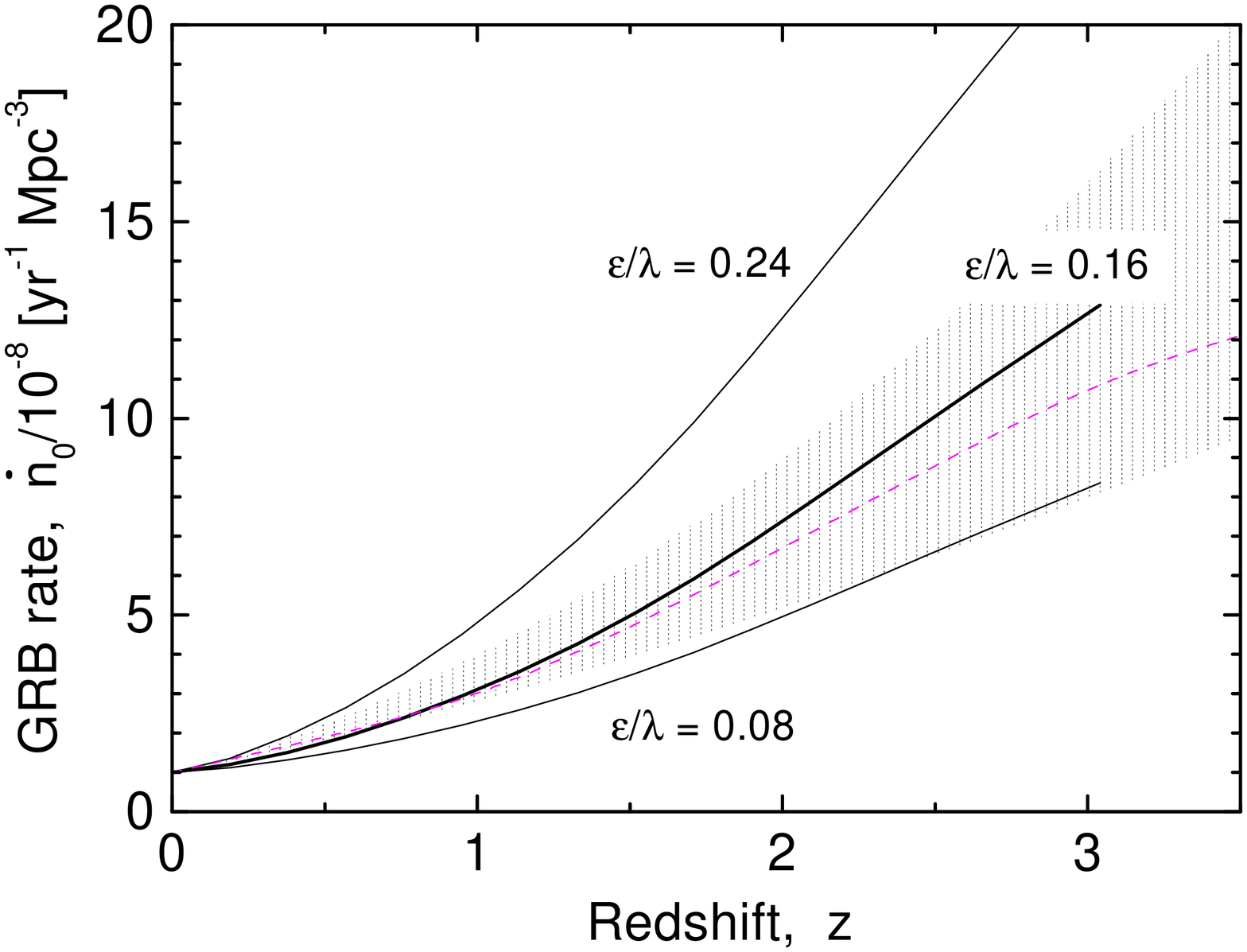,width=3.2in,height=1.9in}}
\vspace{10pt} \caption{The GRB rate as a function of redshift. The
hatched region confines the allowed
GRB redshift distribution from the BATSE data [9].
Three solid lines show 
the GRB rate from fast evolving stellar clusters in distant galactic
nuclei (a constant contribution from the slowly evolving systems
is added) as given by our model 
for the three different values of $\varepsilon/\lambda$ ratio, where 
$\varepsilon$ is the
fraction of galactic nuclei in which SMBHs are formed, and
$\lambda$ is the average quasar luminosity normalized to the 
Eddington luminosity. For comparison, the dashed line shows the rate 
of GRBs from NS coalescences in tight binaries
produced in the distant star-forming galaxies [11].
} \label{myfirstfigure}
\end{figure}

\end{document}